\documentclass[twocolumn,showpacs,prl,preprintnumbers,amsmath,amssymb]{revtex4}
\usepackage{epsfig}
\usepackage{graphicx}
\usepackage{dcolumn}
\usepackage{bm}
\begin{document}
\title{Cooperative Behavior in a Model of Evolutionary Snowdrift Games with $N$-person Interactions}
\author{Da-Fang Zheng}\email{dfzheng@zju.edu.cn}
\affiliation{ Zhejiang Institute of Modern Physics and Department
of Physics, Zhejiang University, Hangzhou 310027, People's
Republic of China}
\author{Haiping Yin}
\author{Chun-Him Chan}
\author{P. M. Hui}
\affiliation{Department of Physics and Institute of
Theoretical Physics, The Chinese University of Hong Kong, Shatin,
Hong Kong, China}

\date{\today}

\begin{abstract}
We propose a model of evolutionary snowdrift game with $N$-person
interactions and study the effects of multi-person interactions on
the emergence of cooperation. An exact $N$-th-order equation for
the equilibrium density of cooperators $x^*$ is derived for a
well-mixed population using the approach of replicator dynamics.
The results show that the extent of cooperation drops with
increasing cost-to-benefit ratio and the number $N$ of interaction
persons in a group, with $x^{*}\sim1/N$ for large $N$.  An
algorithm for numerical simulations is constructed for the model.
The simulation results are in good agreements with theoretical
results of the replicator dynamics.
\end{abstract}

\pacs{87.23.Kg, 02.50.Le, 89.75.Fb}

\maketitle

The theme of how cooperative behavior emerges among competing
entities has attracted the attention of physicists, applied
mathematicians, biologists, and social scientists in recent years
\cite{gardenes,nowak,ohtsuki,pacheco,antal,zhong,wang,santos,hauert,wu}.
There are good reasons that physicists showed much interest in
this problem and have made contributions.  The cooperative
behavior is similar to that in interacting spin systems, and some
important features, e.g., phase transitions and universality which
carry a heavy flavor of statistical physics, have also been
observed in evolutionary models of cooperation with spatial
structures \cite{hauert,szabo}.  Indeed, applying ideas in physics
across different disciplines is a key characteristic of physics in
the new millennium.

A powerful tool to study cooperative phenomena is the theory of
evolutionary games based on such basic models as the prisoner's
dilemma (PD) \cite{axelrod,trivers,hofbauer} and the snowdrift
game (SG) \cite{sugden,smith}. The basic PD is a two-person game
\cite{neumann,rapoport}, in which two players simultaneously
choose one of two possible strategies: to cooperate (C) or to
defect (D). If one plays C and the other plays D, the cooperator
pays a cost of $S=-c$ and the defector receives the highest payoff
$T=b$ ($b>c>0$).  If both play C, each player receives a payoff of
$R=b-c >0$.  If both play D, the payoff is $P=0$. Thus, the PD is
characterized by the ordering $T>R>P>S$ of the payoffs, with
$2R>T+S$. In a single encounter, defection is the better action in
a well-mixed or fully connected population, regardless of the
opponents' decisions.  Allowing for repeated encounters and
evolution of characters could lead to cooperative behavior
\cite{axelrod}. Due to practical difficulties in measuring the
payoffs or even ranking the payoffs accurately
\cite{milinski,turner}, there are serious doubts on taking PD to
be the most suitable model for studying emerging cooperative
phenomena in a competing setting \cite{hauert1}.  The evolutionary
snowdrift game (ESG) has been proposed \cite{hauert1} as an
alternative to PD and has attracted some recent studies
\cite{zhong,wang,santos}.  The basic snowdrift game (SG), which is
equivalent to the hawk-dove or chicken game \cite{sugden,smith},
is again a two-person game.  It is most conveniently described
using the following scenario. Consider two drivers hurrying home
in opposite directions on a road blocked by a snowdrift. Each
driver has two possible actions -- to shovel the snowdrift
(cooperate (C)) or not to do anything (not-to-cooperate or
``defect" (D)).  If the two drivers cooperate, they could be back
home on time and each will get a reward of $b$. Shovelling is a
laborious job with a total cost of $c$.  Thus, each driver gets a
net reward of $R=b-c/2$. If both drivers take action D, they both
get stuck, and each gets a reward of $P=0$. If only one driver
takes action C and shovels the snowdrift, then both drivers can
get through.  The driver taking action D (not to shovel) gets home
without doing anything and hence gets a payoff $T=b$, while the
driver taking action C gets a ``sucker" payoff of $S=b-c$. The SG
refers to the case of $b>c>0$, leading to $T>R>S>P$.  Thus, PD and
SG only differ by the order of $P$ and $S$ in the ranking of the
payoffs.  This seemingly minor difference leads to significant
changes in the cooperative behavior, when evolution of characters
is introduced.  Following replicator dynamics \cite{hofbauer},
there exists a stable state with coexisting cooperators and
defectors in SG for a well-mixed population. More interestingly,
it was found that spatial structures tend to suppress the extent
of cooperation in ESG \cite{hauert1}, in contrast to the common
belief that spatial structure constitutes a favorable ingredient
for cooperation \cite{nowak1,nowak2}.

Most models of evolutionary games proposed so far for studying
cooperative phenomena, including those with competitions among a
group of entities, involve only two-person interactions. In
reality, multi-person interactions are abundant, especially in
biological and social systems.  A representative model is the
so-called public goods game (PGG) \cite{kagel}, for studying group
interactions in experimental economics.  The PGG considers an
interacting group of $N$ agents or players.  Each player either
contributes a public good of value $b$ at a cost $c$ with $0<c<b$,
or does nothing at no cost. With $n$ cooperators in the group, the
total contributions ${\cal{R}} n b$ are divided evenly among all
players in the group, where $\cal{R}$ (${\cal{R}} < N$) is called
the public good multiplier. Thus a cooperator will get a benefit
of ${\cal{R}} nb/N-c$, and a defector gets ${\cal{R}} n b/N$
without doing anything. Obviously, in a one-shot PGG, defectors
outperform cooperators, leading to a Nash equilibrium where all
players are defectors. For $N=2$, PGG reduces to PD and thus PGG
represents a $N$-person prisoner's dilemma game.

Motivated by the recent works on ESG and PGG, we propose and study
a $N$-person interacting model of SG. We refer to our model as the
$N$-person evolutionary snowdrift game (NESG). The key question is
how cooperation is affected by allowing for $N$-person
interactions. The evolution of cooperative behavior in the NESG is
studied analytically within the framework of the replicator
dynamics \cite{hofbauer}.  For arbitrary interacting group size
$N$, an exact $N$-th-order equation for the equilibrium frequency
or fraction of cooperators $x^{*}(r)$ is derived for a well-mixed
population, where $r = c/b$ is a parameter that characterizes the
cost-to-benefit ratio in SG. The equation can be solved
numerically for $x^{*}$ as a function of $r$ for any $N$. As the
size of the interacting group increases, cooperation in NESG
decreases and $x^{*} \sim 1/N$ for large $N$. These results are
checked against results obtained by numerically simulating the
evolutionary dynamics and good agreements are found.

The $N$-person evolutionary snowdrift game is defined as follows.
Consider a system consisting of $N_{all}$ agents. In a $N$-person
game, an agent competes with a group of $N-1$ other agents.
Depending on the situation, the interacting group of $N$ agents
can be chosen at random among the $N_{all}$ agents as in the case
of a well-mixed population or defined by an underlying geometry as
in the case of a regular lattice or other networks.  There is a
task to be done and every agent will get a reward of $b$ if it is
completed by one or more agents within the group.  The total cost
of performing the task is $c$, which could be shared among those
who are willing to cooperate. The payoff of an agent thus depends
on (i) the character of the agent and (ii) the characters of his
$N-1$ competing agents. Here, we will focus on the case of a
well-mixed population.

For an agent of C-character, his payoff depends on the number of
C-character agents in the interacting group including himself. The
C-character agents are those who are willing to share the labor
in completing the task.   If the agent under consideration is the
sole C-character agent in the group, then his payoff is $b-c$. If
there are two C-character agents, then his payoff is $b - c/2$,
and so on. Thus, a C-character agent in a $N$-person snowdrift
game has a payoff of
\begin{equation}\label{equ-Pc}
P_{C} (n) =
    b - \frac cn, \text{\hspace{3ex}for }n \in [1,N],
\end{equation}
where $n$ is the number of C-character agents in the group of $N$
agents including the agent concerned.

For an agent of D-character, his payoff depends on whether there
is a C-character agent in the group. As long as there is one, the
task will be completed and the D-character agent will get a
payoff of $b$ without doing any work. When there is no
C-character in the group, then his payoff vanishes since the
group has $N$ D-character agents and no one is willing to perform
the task. Thus, a D-character agent in a $N$-person snowdrift
game has a payoff of
\begin{equation}\label{equ-Pd}
P_D (n) =
    \begin{cases}
    0 & n=0 \\
    b  & n\in [1, N-1]
    \end{cases}.
\end{equation}
As evolution proceeds in NESG, the numbers of C-character and
D-character agents become time-dependent.

The model is original. It is different from the previous models in
which the payoffs are typically evaluated by summing up the
payoffs of two-player games, for a player competing with a number
of other players.  There are many real-life situations where
pairwise interactions are inapplicable. We give two examples here
where $N$-person interactions are more appropriate.  (i) In a
public construction project such as a bridge, a school or a road
that serves a small remote community, everyone in the neighborhood
will be benefited ($b$) and the cost ($c$) can be shared by those
who are willing to contribute. (ii) A place such as a class room,
a dormitory or a student common room needed to be cleaned
regularly with a labor of cost $c$, and every user will get a
benefit $b$ from the cleanliness.  Certainly, more realistic
modelling will require additional parameters, e.g., more
incentives for carrying out the task in the form of long term
returns.  Here, we study the simplest version as the model can be
treated analytically and thus provides insight into the extent of
cooperation with a function of the parameters $r$ and $N$ in the
model.

The evolutionary behavior in NESG in a well-mixed population is
introduced through the replicator dynamics \cite{hofbauer}. The
frequency of cooperation $x(t) = N_{C}(t)/N_{all}$, where
$N_{C}(t)$ is the number of C-character agents in the population
at time $t$ \cite{zhong,hauert1}.  The time evolution of $x(t)$ is
governed by the following differential equation \cite{hofbauer}
\begin{equation}\label{equ-x}
  \dot{x} = x (f_{C} -\bar{f}),
\end{equation}
where $f_{C}(t)$ ($\bar{f}$) is the instantaneous average fitness
of a C-character agent (the whole population). These quantities
are equivalent to the corresponding average payoffs in the case of
strong coupling \cite{hauert2}.  In the well-mixed case,
interacting groups of $N$ agents are randomly chosen.  The fitness
$f_{C}$, which is in general time dependent, is determined as
follows according to the binomial sampling \cite{hauert2}
\begin{equation}\label{equ-fc}
  f_{C} = \sum_{j=0}^{N-1} \binom{N-1}{j} x^j (1-x)^{N-1-j}
  P_{C}(j+1),
\end{equation}
which takes into account of the various combinations of the
characters of an agent's $N-1$ neighbors. The first three factors
in the sum give the probability of having $(j+1)$ C-character
agents in the group of $N$ agents. Similarly, the instantaneous
average fitness $f_{D}(t)$ or the average payoff of a D-character
agent is given by
\begin{equation}\label{equ-fd}
  f_{D} = \sum_{j=0}^{N-1} \binom{N-1}{j} x^j (1-x)^{N-1-j}
  P_{D}(j).
\end{equation}
These expressions amount to a mean field approach.  In Eq.
(\ref{equ-x}), the dynamics of cooperation is that $x(t)$ will
increase (decrease) if the fitness $f_{C}$ is greater (smaller)
than the instantaneous average fitness $\bar{f}(t)$ of the whole
population.  The latter is defined by
\begin{equation}\label{equ-fit}
\bar{f}(t) = x(t) f_{C}(t) + (1-x(t)) f_{D}(t).
\end{equation}
Substituting Eq.(\ref{equ-fit}) into Eq.(\ref{equ-x}), the
dynamics of $x(t)$ is governed by
\begin{equation}\label{equ-equ}
\dot{x} = x (1-x)(f_{C} - f_{D}).
\end{equation}

Although it is possible to solve the time evolution of $x(t)$, we
will instead focus on the steady state.  After the transient
behavior, the system evolves into a steady state, i.e., the Nash
Equilibrium, in which $\dot{x} = 0$.  It follows from Eq.
(\ref{equ-equ}) that the steady state or equilibrium frequency of
cooperation $x^*$ satisfies
\begin{equation}\label{equ-fcfd}
  f_C (x^*) = f_D (x^*).
\end{equation}
Substituting Eqs. (\ref{equ-Pc}) and (\ref{equ-Pd}) into Eqs.
(\ref{equ-fc}) and ({\ref{equ-fd}) gives $f_{C}$ and $f_{D}$ in
terms of $N$, $b$ and $c$.  Equation (\ref{equ-fcfd}) for $x^{*}$
can then be expressed as
\begin{eqnarray}\label{equ-key}
\sum_{j=1}^{N-1}\frac{1}{j+1}\binom{N-1}{j}(\frac{x^*}{1-x^*})^j=\frac{b-c}{c}.
\end{eqnarray}
Using the identity
\begin{equation}
  (1+y)^N = \sum_{i=0}^{N}\binom{N}{i}y^i,
\end{equation}
we have
\begin{eqnarray}
  \int_0^x (1+y)^N dy & = & \left.\frac{1}{N+1}(1+y)^{N+1}\right|_0^x \nonumber \\
  &=&
  \left.\sum_{i=0}^{N}\binom{N}{i}\frac{y^{i+1}}{i+1}\right|_0^x,
\end{eqnarray}
and thus the relation
\begin{equation}\label{equ-mult}
 \sum_{i=0}^{N}\binom{N}{i}\frac{x^{i+1}}{i+1} = \frac{1}{N+1}[(1+x)^{N+1}
 -1].
\end{equation}
Applying Eq.(\ref{equ-mult}) to Eq.(\ref{equ-key}), we find
\begin{equation}\label{equ-final}
  r (1 - x^*)^N + N x^* (1-x^*)^{N-1} - r =0,
\end{equation}
which is an $N$-th-order equation for $x^{*}(r,N)$ in the steady
state, where $r = c/b$.  Note that the size of the population
$N_{All}$ does not enter, as the analysis assumes an infinite
population following the mean field spirit.

\begin{figure}[h]
\includegraphics[width=0.37\textwidth]{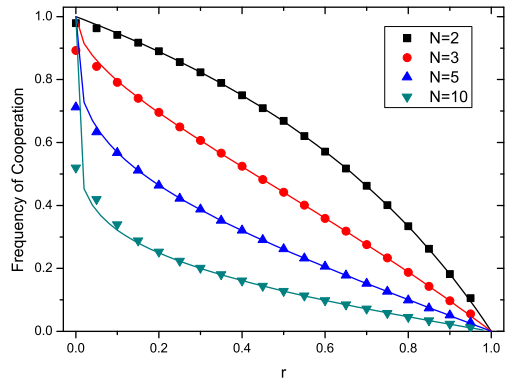}\caption{(Color online) The
equilibrium frequency of cooperation as a function of $r = c/b$,
for $N=2,3,5$, and $10$ in a well-mixed population.  The analytic
results (lines) obtained by solving Eq. (\ref{equ-final}) and the
simulation results (symbols) are in good agreement.  In
simulations, we used $N_{all} = 2000$ and $10^{5}$ time steps.
Each data point is an average over 100 realizations.}\label{fig:1}
\end{figure}
\begin{figure}[h]
\includegraphics[width=0.37\textwidth]{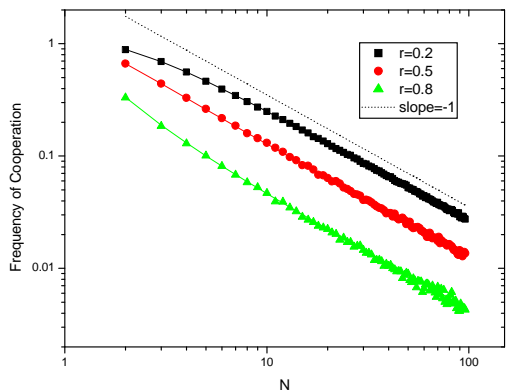}
\caption{(Color online) A log-log plot of the equilibrium
frequency of cooperation as a function of $N$ for $r=0.2, 0.5$,
and $0.8$ in a well-mixed population.  The analytic results
(lines) from Eq. (\ref{equ-final}) and the simulation results
(symbols) are in good agreement.  In simulations, we used $N_{all}
= 5000$ and $10^{7}$ time steps. Every data point is an average
over 10 realizations. A dotted line of slope $-1$ is shown as a
guide to the eye.}\label{fig:2}
\end{figure}

Equation (\ref{equ-final}) can be solved exactly in closed form
for $N\leq 4$. For $N=2$, Eq. (\ref{equ-final}) recovers the
result $x^* = \dfrac{b-c}{b-c/2}$ of the standard two-person
evolutionary SG in a well-mixed population \cite{zhong,hauert1}.
For $N\geq5$, Eq.(\ref{equ-final}) can be solved numerically for
$x^{*}(r,N)$. Figure \ref{fig:1} shows the results (lines) of
$x^{*}(r)$ for $N=2,3,5,10$.  We note that $x^{*}(r)$ decreases as
$r$ increases for arbitrary $N$, with a more rapid drop as $r$
increases for larger values of $N$.  This indicates that the
incentives for being a cooperator drops as $r$ and $N$ increase,
and agents tend to wait for someone else to perform the task and
enjoy a free ride.  For a given $r$, the dependence of $x^{*}$ on
$N$ is shown in Fig. \ref{fig:2} on a log-log scale. The results
(lines) show that $x^{*}$ decreases with increasing $N$, with a
power-law of exponent $-1$ for large $N$. Analytically, the large
$N$ behavior can be extracted by taking the small $x^{*}$ limit of
Eq. (\ref{equ-final}).  We find
\begin{equation}\label{equ-ana-apro}
x^* = \frac{2(1-r)}{(N-1)(2-r)},
\end{equation}
from which $x^{*}\sim1/N$ for large $N$ follows.

As a supplement and to verify the results using the replicator
dynamics, we also perform numerical simulations on NESG. The
algorithm goes as follows. An agent in a total population of
$N_{all}$ agents can take on either the C-character or
D-character. The initial characters of the agents are assigned
randomly. At each time step, an agent $i$ is randomly chosen and a
group of $N-1$ other agents are randomly chosen among the
$N_{all}-1$ agents to compete with $i$. Depending on the character
of agent $i$, his payoff $P_{i}$ is evaluated according to Eq.
(\ref{equ-Pc}) or Eq. (\ref{equ-Pd}). Evolution of character of
agent $i$ is introduced by comparing the payoff with that of
another agent $j$, which is again randomly chosen. For the chosen
agent $j$, he would compete with a randomly chosen group of $N-1$
agents and his payoff is $P_{j}$.  If $P_{i}$ is less than $P_j$,
the character of the agent $i$ will be replaced by that of agent
$j$ with a probability $(P_j-P_i)/b$. If $P_{i} \geq P_{j}$, the
character of agent $i$ remains unchanged. The results from
numerical simulations (symbols in Fig. \ref{fig:1} and Fig.
\ref{fig:2}) are in good agreement with the analytic results based
on the replicator dynamics.  The way of constructing a proper
simulation algorithm will also be useful in studying variations of
the model in which analytic approaches fail.

In summary, we have proposed and studied an evolutionary snowdrift
game with $N$-person interactions. We derived an exact
$N$-th-order equation for the equilibrium frequency $x^{*}(r,N)$
of cooperators in a well-mixed population using the approach of
replicator dynamics. The results show that the level of
cooperation lowers as $r$ increases.  For fixed $r$, $x^{*}$ drops
with the number $N$ of interaction persons in a group and takes on
$x^{*}\sim1/N$ for large $N$.  We also constructed a numerical
algorithm to simulate the model.  Simulation data are in good
agreement with the analytic results of the replicator dynamics.
Further extension of NESG to include the effects of spatial
structures such as regular lattices and complex networks will be
interesting.

This work was supported in part by the National Natural Science
Foundation of China under Grant No. 70471081 (DFZ), and by the
Research Grants Council of the Hong Kong SAR Government under
Grant No. CUHK-401005 (PMH).}

\end{document}